# Research on the multi-stage impact of digital economy on rural revitalization in Hainan Province based on GPM model

**Wenbo Lyu** [a#]

[a] *Saxo Fintech Business School, University of Sanya, Sanya, 572000, China*
**Corresponding author.*
*Tel/Fax: +86 18617571856; E-mail: wenbolv@sanyau.edu.cn; Orcid ID：0009-0005-4329-5616*

**Abstract:** In the context of the new era, the digital economy has become a new engine for global economic growth. As the southernmost province in China, Hainan Province has unique geographical locations and resource advantages. The rapid development of the digital economy has had a profound impact on the implementation of the rural revitalization strategy. Based on this, this study takes Hainan Province as the research object to deeply explore the impact of digital economic development on rural revitalization. The study collected panel data from 2003 to 2022 to construct an evaluation index system for the digital economy and rural revitalization and used panel regression analysis and other methods to explore the promotion effect of the digital economy on rural revitalization. Research results show that the digital economy has a significant positive impact on rural revitalization, and this impact increases as the level of fiscal expenditure increases. The issuance of digital RMB has further exerted a regulatory effect and promoted the development of the digital economy and the process of rural revitalization. At the same time, the establishment of the Hainan Free Trade Port has also played a positive role in promoting the development of the digital economy and rural revitalization. In the prediction of the optimal strategy for rural revitalization based on the development levels of the primary, secondary, and tertiary industries (Rate1, Rate2, and Rate3), it was found that rate1 can encourage Hainan Province to implement digital economic innovation, encourage rate3 to implement promotion behaviors, and increase rate2 can At the level of sustainable development when rate3 promotes rate2's digital economic innovation behavior, it can standardize rate2's production behavior to the greatest extent, accelerate the faster application of the digital economy to the rural revitalization industry, and promote the technological advancement of enterprises. This study not only provides theoretical support and practical reference for Hainan Province to explore effective ways for the digital economy to promote rural revitalization but also provides reference for other regions across the country. The research conclusion emphasizes the importance of fiscal expenditure, digital RMB, and free trade port policies in promoting the development of the digital economy and rural revitalization, providing useful enlightenment for policymakers.

## 1 Introduction

With the rapid development of global digital technology, the digital economy has become a new engine driving world economic growth. The Chinese government attaches increasing importance to the digital economy and has proposed a series of strategic plans and policy measures at the national level. The report of the 20th National Congress of the Communist Party of China clearly stated that it is necessary to accelerate the construction of a cyber power and a digital China and promote the deep integration of the digital economy and the real economy. In February 2023, the State Council issued Document No. 1, which proposed more new requirements for agricultural digitalization for rural revitalization. The document pointed out that the "Digital Rural Development Strategy Outline" and "Digital Rural Development Action Plan (2022-2025)" should be implemented in depth to empower digitalization. It can promote the development of rural industries, rural construction, and rural governance, promote the development of agricultural and rural modernization as a whole, promote the common prosperity of rural farmers, and push the construction of digital China to a new level. This strategic orientation provides macro policy support and theoretical guidance for Hainan Province to explore the digital economy to help rural revitalization (Chen Yang, 2022).

As the southernmost province in China, Hainan Province has unique geographical locations and resource advantages and has also become a key support target for national rural revitalization. From the report of the 19th National Congress of the Communist Party of China proposing to explore the construction of a free trade port to General Secretary Xi Jinping announcing support for the construction of a free trade pilot zone in Hainan at the conference celebrating the 30th anniversary of the establishment of Hainan Province as a special economic zone to the Central Committee of the Communist Party of China and the State Council issuing the "Hainan "Overall Plan for the Construction of the Free Trade Port", this series of important decisions and deployments provide clear policy guidance and institutional guarantees for the construction of the Hainan Free Trade Port (Zhang Lin, 2023). In the context of the construction of the free trade port, Hainan Province is actively exploring new paths for digital economic development to promote regional economic growth and industrial structure optimization. Since 2018, Hainan's digital economy, dominated by the Internet industry, has continued to grow rapidly. Hainan Province The core industries of the digital economy will achieve revenue of 127.96 billion yuan in 2022, accounting for 7.5% of GDP and contributing 47% to high-tech industries. , At the same time, Hainan, as the only region in the country that has piloted digital renminbi across the province, has many applications in both C-side and B-side scenarios. By 2022, Hainan's digital renminbi will support 50,000 merchants and open personal wallets. 640,000 (Chen Xinxin, 2023). However, the rural revitalization and development of Hainan Province also face many challenges, such as lagging infrastructure construction, single industrial structure, and lack of human resources. Therefore, how to use the digital economy to promote rural revitalization and achieve coordinated development of urban and rural areas has become an important issue facing the development of the digital economy in Hainan Province.

In the 1990s, Don Tapscott proposed the early concept of the digital economy. Since then, the digital economy concept has gradually been regarded as the "national information highway." At the micro level, the digital economy is changing traditional rural production methods and lifestyles. Through the application of technologies such as the Internet, big data, and artificial intelligence, agricultural production efficiency has been improved, agricultural product sales channels have been expanded, and emerging industries such as rural tourism and e-commerce have flourished. These changes have brought more development opportunities and employment opportunities to rural areas and also improved the living standards and happiness of rural residents. Based on this, this study collects 20-year panel data from 2003 to 2022, aiming to deeply explore the multi-stage impact of Hainan Province's digital economy on rural revitalization. First, by constructing the digital economy development level indicator system and the rural revitalization development indicator system, the digital economy index (DEI) and the rural revitalization index (Rural) are calculated to effectively measure the explanatory variable digital economy and the explained variable rural revitalization in this article, and select the aging population. The variables (Age), household consumption level (CPI), trade openness (Trade), economic level (GDP), primary, secondary, and tertiary industry development levels (Rate1, Rate2, Rate3) are used as control variables, and fiscal expenditure is Level (Tr) is used as the threshold variable, and panel regression is used to analyze the promotion effect of the digital economy on rural revitalization. The digital renminbi (digital currency issued by the central bank) is introduced as an adjusting variable to explore the important regulation that the digital renminbi plays in the digital economy and rural revitalization. effect; secondly, a Game-fusion Parzen-window model (GPM) is established to predict and analyze the overall optimal solution of GDP/CPI equilibrium and coordinated equilibrium of the three major industries, and eliminate the local optimal solution of short-term growth in the real estate industry. Excellent solution; based on the empirical results, policy suggestions and measures to promote the deep integration of digital economy and rural revitalization in Hainan Province are put forward to provide theoretical support and practical reference for Hainan Province and other regions across the country to explore effective paths for the digital economy to assist rural revitalization.

## 2 Literature review

### 2.1 Related research on the digital economy

So far, all walks of life have their definitions of the connotation and definition of the digital economy, and there is no unified standard answer. The digital economy, a concept that originated in the 1990s, has now become the core driving force of the global economy. It is based on information communication and digital technology and covers a wide range of activities from digital industrialization to industrial digitization. Digital industrialization mainly includes the development of information technology industries, such as electronic information business, communication business, and software business; while industrial digitization focuses on Focus on using new information technology to digitally transform traditional industries throughout the entire process.

There are various methods of measuring the digital economy. The "Measuring the Information Society Report" and the IDI index released by the United Nations International Telecommunications Union are widely adopted and are mainly measured from three aspects: ICT access, usage, and skills (Zhou Jian, 2023). In addition, various countries and organizations have also proposed different measurement methods, such as CCID Consulting's DEDI Index (Qin Hailin, 2020), the Organization for Economic Cooperation and Development's Social Empowerment Indicator, etc. (Huang Zhongjing, 2020). These methods have their advantages and disadvantages in terms of data sources, stability, and comprehensiveness.

The impact of the digital economy on the economy and society is far-reaching, with both positive and potential negative effects. On the positive side, the digital economy injects new vitality into economic development by promoting market innovation, enhancing the economy's ability to resist risks, and improving government social governance capabilities. However, the digital economy also brings some challenges. On the one hand, it may lead to the "digital divide" (Zhang Lun, 2013) and the "Matthew Effect" (Qin Hongwei, Dong Jing, 2021), making some groups unable to cope with the development of the digital economy. at a disadvantage. On the other hand, the development of the digital economy poses new challenges to government supervision, such as information security and privacy protection.

## 2.2 Related research on rural revitalization

Rural revitalization is a global issue that has not only received widespread attention in China but has also triggered in-depth research and practice abroad. In China, General Secretary Xi Jinping proposed the rural revitalization strategy for the first time in the report of the 19th National Congress of the Communist Party of China in 2017, emphasizing that the "agriculture, rural areas, and farmers" issues are fundamental to the national economy and people's livelihood, and must make solving the "agriculture, rural areas, and farmers" issues a top priority in the work of the whole party. (Peng Zhaorong, 2023). This strategy covers five aspects: industrial prosperity, ecological livability, civilized rural customs, effective governance, and affluent life. It provides a direction for the comprehensive development of rural areas from the economic, ecological, cultural, political, and social levels. In foreign countries, although there is no rural revitalization strategy the same as that in China, governments of various countries are also working hard to promote the comprehensive development of rural economic ecology. For example, Japan encourages high-tech enterprises to leave big cities and combine them with the natural environment of rural areas to form "Green Silicon Valley". The United States supports rural economic development and job creation through rural economic development loans and grant programs.

Domestic and foreign scholars have conducted a large number of studies on measuring the level of rural revitalization. In China, the early socialist new rural evaluation indicator system provided the basis for the current rural revitalization indicators. With the progress of rural areas and the new requirements of policies, the measurement standards of rural revitalization are also constantly improving. Many scholars construct an evaluation index system for rural revitalization from five aspects: industry, ecology, culture, governance, and life. At the same time, some scholars analyze and summarize from other perspectives, such as the efficiency, quality, and modernization of agriculture and agricultural products (Zhang Ph. , 2022; Luo Haichao, 2023; Li Lu, 2019).

When implementing rural revitalization strategies, countries are faced with the problem of how to provide policy support based on local conditions. For example, the Canadian government divides regions into different types based on population size and urban radiation intensity and provides policy support with different weights based on the main rural economic types.

## 2.3 Related research on the impact of the digital economy on rural revitalization

Globally, although there is no rural revitalization strategy the same as that in China, the impact of rural economic development and the digital economy on rural revitalization has received widespread attention. Reto Bürgin's research reveals the two-way effect of digitalization on rural revitalization. The digital economy has brought countless new opportunities to large enterprises and government agencies, but for rural small businesses, how to gain a foothold in the digital economy is a huge challenge. Grimes S.'s survey of more than 600 districts in six European countries showed that even when telecommunications and information technology infrastructure is well-established, there are still huge differences in the ability of rural areas to use these technologies in competitive means. This shows that, in addition to Infrastructure high-quality talents are also the keys to promoting the digital economy and supporting rural development.

The relationship between the digital economy and rural development has also been deeply discussed in China. In the four dimensions of technology, factors, products, and institutions, the digital economy can provide strong assistance for rural revitalization (Zhu Xiaotong, 2023). The inclusiveness and spillover nature of digital technology can break the limitations of traditional industries, give birth to new business formats and industrial models, and help comprehensively upgrade rural areas and increase farmers' income (Chai Hailiang, 2022). As a new element, data not only drives green and high-quality rural development but also promotes the flow of rural natural resources. The digital economy can also quickly provide unique and high-quality products and services to meet the diverse needs of consumers while spreading rural specialty products and culture to the world (Zhang Jiashi, 2023).

However, the integration of the digital economy and rural revitalization also faces many challenges. Wen Tao pointed out that insufficient infrastructure in rural areas, poor data circulation, lack of digital economy talents, and lack of relevant standards, laws, and regulations are the main bottlenecks restricting the integration of the two. Xie Wenshuai proposed that the platform operation of economic organizations and the digital governance of governments should work together to promote coordinated changes in rural economic structure, production relations, and superstructure. Wang Weiling emphasized the government's top-level design and long-term planning role in the development of the digital economy and built a complete rural digital economic ecosystem to help rural revitalization.

### 2.4 Research review

Through a literature review, it was found that research on the digital economy and rural revitalization has achieved remarkable results both at home and abroad. Although rural digitalization in my country started later than abroad, the integration of the digital economy and rural industries is regarded as an important way to revitalize rural areas. Currently, digital economy research mostly focuses on digital inclusive finance, while rural revitalization mainly focuses on narrowing the urban-rural income gap. However, the digital economy in the new era involves industrial digitization, digital industrialization, and other fields. Rural revitalization also requires comprehensive promotion of agricultural and rural modernization from multiple dimensions. Considering the spatial effects of the digital economy and its obvious spillover effects, relevant research is still insufficient. Given this, this article selects panel data from Hainan Province in my country from 2003 to 2022 to construct an evaluation index system for the development level of the digital economy and rural revitalization development and uses methods such as panel models, game models, and Bayesian models to deeply explore how the digital economy develops. influence rural revitalization. For Hainan Province, this research has important practical significance and helps guide the local area on how to use the digital economy to promote the modernization process of agriculture and rural areas and achieve multi-dimensional comprehensive revitalization.

## 3 Research methods and materials

### 3.1 Constructing an index system based on the entropy weight TOPSIS method

(1) Indicator selection and measurement of digital economy development level

The digital economy development level evaluation index system constructed in this article comprehensively reflects the digital economy development level. The first-level indicators cover the foundation of the digital economy, the integration of digital and industry, and the use of digital finance. The secondary indicators select 6 representative indicators, including mobile phone penetration rate, number of 4G base stations per capita, etc., to reflect the basic role of the digital economy; the proportion of telecommunications business in GDP and other measures measure the degree of digital industrialization; the proportion of R&D expenditures in GDP measures the regional commitment to R&D Pay attention; the digital finance digitization degree index shows the deep integration of digital finance and digital economy.

Table 3-1 Digital economy indicator system

| Overall indicator | First level indicator | Secondary indicators | direction |
|---|---|---|---|
| Digital economy development level | Digital economic development carrier | Mobile phone penetration rate (units/100 people) | + |
| | | Number of 4G mobile phone base stations per capita (units/100 people) | + |
| | Digital industrialization level | Telecom business revenue as a share of GDP (%) | + |
| | | The proportion of computer services and software employees (%) | + |
| | Digital economic development environment | R&D expenditure as a share of GDP (%) | + |
| | Industrial digitalization level | Degree of digitalization of Peking University's digital finance | + |

Data source: Hainan Province Statistical Yearbook 2003-2022

In the comprehensive calculation of the digital economy development level index, to enable the indicator data to avoid the influence of excessive units and values to a certain extent, the obtained data needs to be processed dimensionless. The basic idea of dimensionless processing is to process the original data through mathematical calculations and convert the original data of different dimensions or units from absolute size to relative size. This paper uses the extreme value method to dimensionally process the data from both positive and negative directions.

Positive indicators: $X'_{ij} = \frac{X_{ij}-\min\{X_j\}}{\max\{X_j\}-\min\{X_j\}}$(3.1)

Negative indicator: $X'_{ij} = \frac{\max\{X_j\}-X_{ij}}{\max\{X_j\}-\min\{X_j\}}$(3.2)

Among them, $X_{ij}$the 6 indicator values $\min\{X_i\}$expressing the arrangement of the year and item $j$in vector form are the smallest in all years among the 6 secondary indicators in vector form and are the smallest indicator values in all years $\max\{X_j\}$among the 6 secondary indicators in vector form. The largest indicator value in. $X$It cannot be 0 in subsequent logarithmic calculations, so when the dimensionless value is 0, it is assigned a value of 0.00001.

This article chooses the entropy weight method as the calculation method of weight. First, calculate the proportion of the index value of the first year:$j$

$$w_{ij} = \frac{x'_{ij}}{\sum_{i=1}^{m} X_{ij}} (3.3)$$

Among them, is the total number of years to be calculated, $m$ =20 in this article. The next step is to calculate the information entropy and information entropy redundancy of the indicator.

Information entropy:

$$e_j = -\frac{1}{\ln m} \sum_{i=1}^{m} \left( w_{ij} \times \ln w_{ij} \right) (3.4)$$

The information entropy redundancy is:

$$d_j = 1 - e_j (3.5)$$

Finally, the weight is calculated based on the information entropy and its redundancy:

$$w_j = \frac{d_j}{\sum_{i=1}^{n} d_j} (3.6)$$

Among them, represents the number of indicators, n =20 in this article.

After the weight is obtained through the calculation of the dimension index, the score of the second-level indicator such as mobile phone penetration rate in the year is calculated. The score is obtained by multiplying the weight and the dimensionless value of the indicator $X'_{ij}$to to obtain the score of the indicator in the year. i:

$$S_{ij} = w_j \times X'_{ij} (3.7)$$

To calculate the development level of the digital economy, after obtaining the scores of each indicator in the year, the digital economy development level index is obtained by adding up the sum (as shown in Table 3-2):

$$\text{Score}_i = \sum_{j=1}^{n} S_{ij} (3.8)$$

Table 3-2 Digital economy index by region from 2003 to 2022

| City | 2003 | 2004 | 2005 | 2006 | 2007 | 2008 | 2009 | 2010 | 2011 | 2012 |
|---|---|---|---|---|---|---|---|---|---|---|
| Haikou | 0.048 | 0.070 | 0.043 | 0.005 | 0.098 | 0.077 | 0.066 | 0.030 | 0.048 | 0.070 |
| Sanya City | 0.086 | 0.085 | 0.069 | 0.055 | 0.079 | 0.058 | 0.047 | 0.097 | 0.086 | 0.091 |
| Sansha City | 0.003 | 0.033 | 0.087 | 0.068 | 0.078 | 0.048 | 0.003 | 0.080 | 0.003 | 0.033 |
| Danzhou City | 0.009 | 0.035 | 0.007 | 0.095 | 0.098 | 0.056 | 0.057 | 0.064 | 0.009 | 0.035 |

| City | 2013 | 2014 | 2015 | 2016 | 2017 | 2018 | 2019 | 2020 | 2021 | 2022 |
|---|---|---|---|---|---|---|---|---|---|---|
| Haikou | 0.094 | 0.117 | 0.207 | 0.155 | 0.187 | 0.299 | 0.366 | 0.410 | 0.455 | 0.469 |
| Sanya City | 0.104 | 0.127 | 0.199 | 0.181 | 0.210 | 0.294 | 0.351 | 0.38 9 | 0.399 | 0.412 |
| Sansha City | 0.095 | 0.113 | 0.207 | 0.178 | 0.197 | 0.283 | 0.330 | 0.3 3 6 | 0.352 | 0.389 |
| Danzhou City | 0.061 | 0.080 | 0.149 | 0.121 | 0.139 | 0.237 | 0.317 | 0.3 3 1 | 0.541 | 0.366 |

Data source: Hainan Province Statistical Yearbook 2003-2022

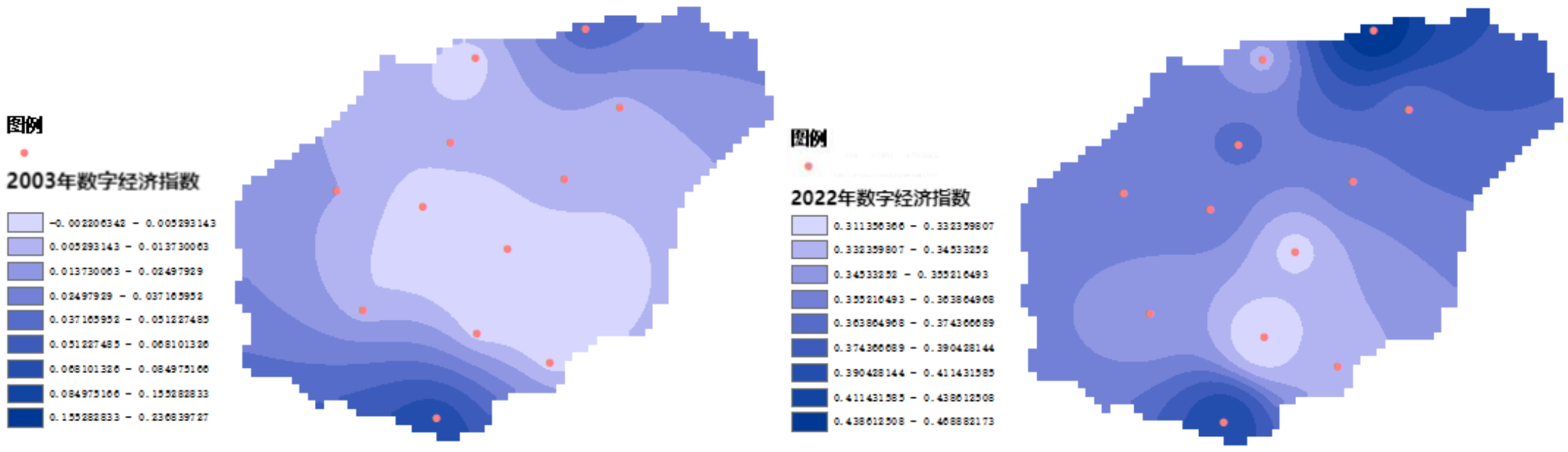

Note: Made in ARCGIS10.0, the affiliated islands are not shown.

Figure 3-1 (a) Distribution of digital economy index in Hainan Province in 2003 Figure 3-1 (b) Distribution of digital economy index in Hainan Province in 2022

Figure 3-1 Spatial distribution of the inverse distance of digital economy index in Hainan Province

(2) Indicator selection and measurement of rural revitalization development level

This article draws on the research results of Zhao Lifang (2021) and others on rural revitalization and uses 13 indicators such as the Engel Index of rural residents to construct a measurement of rural revitalization from the five dimensions of industrial prosperity, ecological livability, rural civilization, effective governance, and affluent life. indicator system. The specific indicators are shown in Table 3-3:

Table 3-3 Rural revitalization development level evaluation index system

| main indicator | First level indicator | Secondary indicators | direction |
|---|---|---|---|
| Rural Revitalization Index | live a prosperous life | Engel Index of Rural Residents | - |
| | | Salary income as a share of total income | + |
| | | Rural per capita disposable income | + |
| | Ecological and livable | Pesticide usage (tons/hectare) | + |
| | | Hygienic toilet penetration rate (%) | + |
| | Rural customs and civilization | Rural radio programs (%) | + |
| | | Rural TV programs (%) | + |
| | | Average consumption expenditure on culture, education, and entertainment per person of rural households (yuan/person) | + |
| | Industry is booming | Consumption level of rural residents (yuan/person) | + |
| | | Total agricultural machinery power per capita (kW/person) | + |
| | Effective Governance | Number of cars owned by rural households per 100 households (Taiwan/100 households) | + |
| | | Income ratio of rural residents to urban residents (%) | + |
| | | The proportion of rural residents receiving minimum subsistence allowance ( % ) | - |

Data source: Hainan Province Statistical Yearbook 2003-2022

The calculation method of the rural revitalization index here is the same as above, and will not be repeated. The rural revitalization index of each province from 2003 to 2022 is obtained.

Table 3-4 Rural revitalization index by region from 2003 to 2022

| City | 2003 | 2004 | 2005 | 2006 | 2007 | 2008 | 2009 | 2010 | 2011 | 2012 |
|---|---|---|---|---|---|---|---|---|---|---|
| Haikou | 0.323 | 0.379 | 0.415 | 0.429 | 0.454 | 0.473 | 0.402 | 0.481 | 0.484 | 0..485 |
| Sanya City | 0.317 | 0.370 | 0.393 | 0.415 | 0.432 | 0.446 | 0.476 | 0.481 | 0.491 | 0.512 |
| Sansha City | 0.354 | 0.394 | 0.412 | 0.334 | 0.367 | 0.398 | 0.330 | 0.336 | 0.362 | 0.387 |
| Danzhou City | 0.340 | 0.395 | 0.319 | 0.347 | 0.377 | 0.301 | 0.351 | 0.347 | 0.333 | 0.0.327 |
| City | 2013 | 2014 | 2015 | 2016 | 2017 | 2018 | 2019 | 2020 | 2021 | 2022 |
| Haikou | 0.513 | 0.562 | 0.587 | 0.622 | 0.653 | 0.708 | 0.730 | 0.682 | 0.734 | 0.740 |
| Sanya City | 0.552 | 0.582 | 0.620 | 0.645 | 0.674 | 0.659 | 0.682 | 0.667 | 0.552 | 0.560 |
| Sansha City | 0.452 | 0.469 | 0.499 | 0.497 | 0.520 | 0.560 | 0.584 | 0.592 | 0.452 | 0.442 |
| Danzhou City | 0.324 | 0.365 | 0.386 | 0.378 | 0.383 | 0.405 | 0.415 | 0.411 | 0.324 | 0.329 |

Data source: Hainan Province Statistical Yearbook 2003-2022

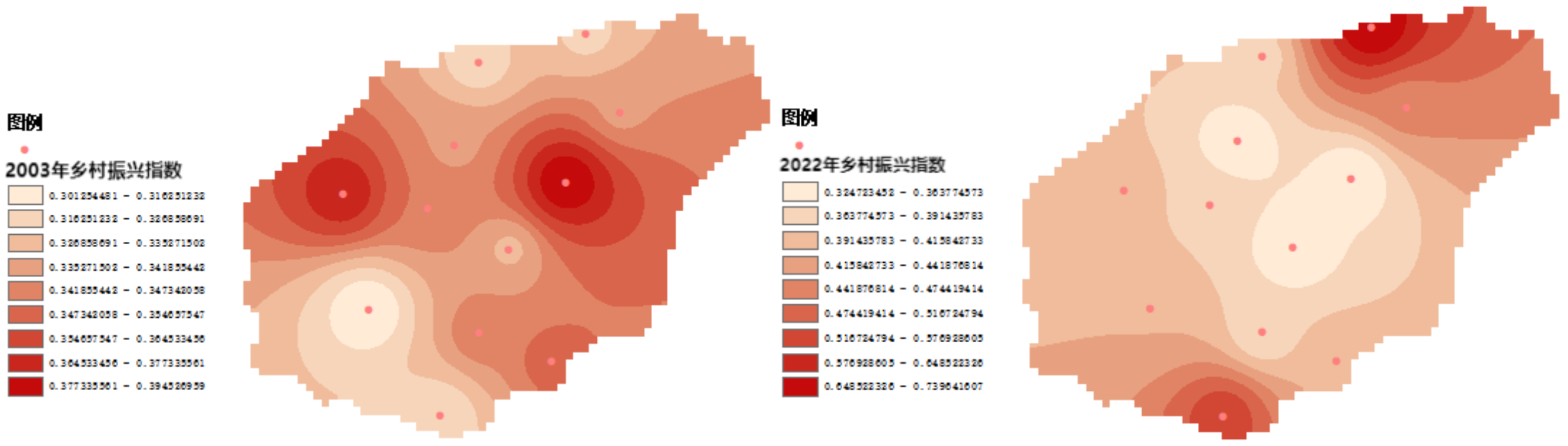

Note: Made in ARCGIS10.0, the affiliated islands are not shown.
Figure 3-2 (a) Distribution of Rural Revitalization Index in Hainan Province in 2003 Figure 3-2 (b) Distribution of Rural Revitalization Index in Hainan Province in 2022

Figure 3-2 Spatial distribution of the inverse distance of rural revitalization index in Hainan Province

## 3.2 Threshold effect test based on spatial panel Durbin model

The spatial panel models selected in this article are mainly divided into types: spatial lag model (SAR), spatial error model (SEM), and spatial Durbin model (SDM). Among them, SAR mainly focuses on the spatial lag term of the dependent variable, while SEM focuses on random The spatial lag term of the interference term, while SDM combines the advantages of the first two and takes into account the spillover characteristics of the spatial lag term of the dependent variable and independent variable.

Therefore, when constructing the spatial econometric model, this paper selects the spatial Durbin model (SDM) of the fixed effects model as the optimal choice based on the views of Lian Yujun (2014). The fixed effect spatial Durbin model of the impact of the digital economy on rural revitalization is set as follows:

$$\text{Rural}_{\text{it}} = \alpha_0 + \rho W\text{Rural}_{it} + \varphi_1 WDEI_{it} + \alpha_1 DEI_{it} + \varphi_2 WX_{it} + \alpha_2 X_{it} + \mu_i + \delta_t + \varepsilon_{it} \quad (3.9)$$

Among them, $\text{DEI}_{\text{it}}$Rural represents the digital economy index, and Rural represents the rural promotion index, which X represents 7 control variables: population aging (Age), household consumption level (CPI), trade openness (Trade), economic level (GDP), first, The development level of the secondary and tertiary industries (Rate1, Rate2, Rate3). ρ is the spatial lag coefficient, $\alpha_1$ is the regression coefficient of the digital economy index, $\alpha_2$ is the regression coefficient of the control variables population aging, resident consumption level, trade openness, economic level, and primary, secondary, and tertiary industry development levels. Vector, $\varphi_1$ is the lag term regression coefficient of the digital economy development level, $\varphi_2$ is the vector of the lag term regression coefficient of the four control variables, $\mu_i$ is the individual effect of the model, $\delta_t$ is the time effect, $\varepsilon_{it}$ is the random error term.

### 3.3 Rural revitalization balanced development strategy based on the game and Bayesian fusion model

The Game and Bayesian Fusion Model (GPM) is a machine learning analysis tool that combines the principles of game theory and Bayesian statistics and is suitable for modeling and prediction of complex economic systems. When discussing the coordination and equilibrium of the three major industries, GPM can be used to effectively analyze the strategic interaction between various economic entities and the decision-making process under uncertainty. Considering the particularity of the real estate industry and its important position in the Chinese economy, eliminating the local optimal solution for its short-term growth is crucial to understanding the overall economic equilibrium. The real estate industry is often affected by multiple factors such as policy, credit, and land supply. Its rapid growth in the short term may mask structural problems in other industries, thereby affecting the judgment of the overall economic health. To achieve the optimal strategy for the balanced development of rural revitalization under this condition, this article designed the following game and Bayesian fusion model (GPM).

Solving the core problem of fusion is equivalent to solving $P(S_1, S_2, \cdots, and\ S_k \mid F_i)$. It can be regarded as the estimation problem of the probability density function. The solution method can be divided into a parametric estimation method and a non-parametric estimation method. Since the distribution form of the probability density is usually unknown during the fusion process, and multimodal functional forms often appear, the Pazen window method of non-parametric estimation is used in the algorithm research of this article to gradually construct and approximate the target probability density distribution.

In the basic Pazen window estimation method, the window function is defined as $\varphi(\mu)$, and satisfies $\varphi(\mu^{\mu}) \geq 0$and satisfies $\int \varphi(\mu)\mathrm{d}\mu = 1$, then the basic formula of Parzen window estimation is:

$$P_N(x) = \frac{1}{N}\sum_{i=1}^{N}\frac{1}{V_N}\varphi\left(\frac{x-x_i}{h_N}\right)(3.10)$$

Assume that all possible decision-making results of the fusion center during the interaction process are $F = \{F_1, F_2, \cdots, F_m\}$. During the fusion process, $\sigma$the information that affects the fusion decision-making among the information collected at all times is. In the estimation $s_\sigma = (s_{a1}, s_\alpha, \cdots, s_{ak})$of C $P(S_2 S_3, ..., S_k \mid F_i)$, use the evaluation class at the current moment $F_i$to obtain the current The information is combined with the acquired preprocessed data $F_i$of the $\sigma -$ 1past classes to form the current $F_i$class sample matrix, and $\sigma$the class conditional probability density function of $P(S_1, S_2, \cdots, S_k \mid F_i)$the moment class is constructed based on this sample matrix $F_i$.

the estimated $P(S_1, S_2, \cdots, S_k \mid F_i)$value, and its product with the payment value (predicted utility) are calculated, and the largest one is used as the final result of the fusion. The specific fusion algorithm based on the Parzen window is described as follows:

(1) Set the loop variable $i = 1$, period calculation result $(l, \text{result }) = (1{,}0)$; (2) Concentrate the samples of the $S_\sigma = (sa1, sa2, \cdots, ss_\alpha)$existing classes at the $F_i$added time $1 \sim \sigma - 1$to construct $F_i$the sample matrix $(s_{ij})\alpha$*k of the class at the current time; (3) Use the Parzen window formula to find $F_i$ the class The distribution density of each component $P_\sigma(s_j \mid F_i) = \frac{1}{\sigma}\sum_{i=1}^{\sigma}\frac{1}{h_\sigma}\varphi\left(\frac{s_{aj}-s_{ii}}{h_\sigma}\right)$, $j = 1,2, ..., \sigma$; (4) Because each data information is independently observed, calculate $F_i$ the conditional probability density of the current class $P_\sigma(s_1, s_2, \cdots, s_n \mid F_i) = \prod_{j=1}^{n} P_\sigma(x_j \mid F_i)$; (5) Whether the calculation $\mu = P(S_1, S_2, \cdots, S_k \mid F_i) \times I_i$is greater than result, if so, the result of this calculation will be saved $(t, \text{result }) = (i, \mu)$; (6) Loop variable $i = {}_i + 1$, and Determine whether there is $i > m$, if so, $F_i$output and the algorithm ends; otherwise, go to step (2).

## 3.4 Variable selection and data sources

The explained variable in this article is the rural revitalization index (Rural) measured previously. The core explanatory variable of this article is the digital economy index (DEI) measured previously. Seven variables including population aging (Age), household consumption level (CPI), trade openness (Trade), economic level (GDP), primary, secondary, and tertiary industry development levels (Rate1, Rate2, Rate3) are used as controls Variables, the fiscal expenditure level (Tr) is used as the threshold variable, and the digital renminbi is selected as the adjusting variable. Among them, population aging is measured by the old-age dependency ratio, trade openness is measured by the proportion of total imports and exports in GDP and the economic level is measured by national product. The total value of GDP is measured by the logarithm, fiscal expenditure is measured by the proportion of general public budget expenditure in GDP, and the development level of the primary, secondary, and tertiary industries is measured by the proportion of primary, secondary, and tertiary industries respectively.

This article selects Hainan Province in my country as the research object. The data also includes four prefecture-level cities, namely Haikou City, Sanya City, Sansha City, and Danzhou City. The data selects relevant data from 2003 to 2022. The data comes from "The China Statistical Yearbook, the National Bureau of Statistics official website, EPS database.

# 4 Result analysis

## 4.1 Analysis of Panel Regression Results

This article first estimates the panel fixed effects method on the panel econometric model to observe the impact of the digital economy on rural revitalization. The model estimation results obtained using SPSS27.0 software are shown in Table 4-1. This paper compares the spatial Durbin model under panel data fixed effects and double fixed effects, and the model fitting goodness is 0.894 and 0.843 respectively.

Table 4-1 Regression results of panel data fixed effects and spatial Durbin model fixed effects

| Interpreted variable: Rural | FE | SDM(both) |
|---|---|---|
| DEI | 0.157***<br>(4.61) | 0.256***<br>(3.92) |
| Age | 0.114<br>(0.97) | -0.09<br>(-0.87) |
| CPI | -0.281*<br>(-1.94) | 0.171**<br>(1.31) |
| Trade | 0.121***<br>(7.62) | 0.061***<br>(3.9) |
| GDP | 0.006***<br>(10.03) | -0.104***<br>(-2.68) |
| Rate1 | 0.353*<br>(1.84) | 0.001**<br>(0.56) |
| Rate2 | 0.458*<br>(1.5) | 0.233**<br>(3.68) |
| Rate3 | 0.676*<br>(1.78) | 0.073**<br>(1.06) |
| rho | -- | 0.354***<br>(3.38) |
| sigma2_e | -- | 0.0003***<br>(10.84) |
| r2 | 0.894 | 0.843 |
| N | 80 | 80 |

Note: ***, **, and * indicate that the significance levels of 1%, 5%, and 10% have been passed respectively. The FE brackets are the t value, the SDM brackets are the z value, and -- indicates that there is no data for this item.

According to the regression results in Table 4-1 above, it can be seen that the digital economy can have a positive and significant impact on rural revitalization. Among them, population aging (Age) has no significant impact, and household consumption level (CPI), trade openness (Trade), economic level (GDP), and primary, secondary, and tertiary industry development levels (Rate1, Rate2, Rate3) are all There is a significant influence relationship.

## 4.2 Threshold effect analysis

Since the population aging (Age) in the control variable is not significant, this variable is eliminated, fiscal expenditure is added as a threshold variable, and threshold effect regression analysis is performed, as shown in Table 4-2:

Table 4-2 Threshold model regression results on the impact of the digital economy on rural revitalization using fiscal expenditure as the threshold

| variable | Parameter Estimation | t | P>t |
|---|---|---|---|
| CPI | 0.2122** | 5.31 | 0.000 |
| Trade | 0.3081** | 2.25 | 0.025 |
| lnGDP | 0.1513*** | 8.42 | 0.000 |
| Rate1 | 0.0112*** | 4.17 | 0.000 |
| Rate2 | 0.0314*** | 2.13 | 0.000 |
| Rate3 | 0.0052*** | 11.80 | 0.000 |
| DEI (Tt≤0.2818) | 0.0568* | 1.81 | 0.058 |
| DEI (Tt>0.2818) | 0.1785*** | 5.55 | 0.000 |
| _cons | -1.538*** | 3.04 | 0.000 |
| R2 | 0.813 | | |

Note: ***, **, and * indicate significance at 1%, 5%, and 10% levels respectively.

According to the data in Table 4-2, it can be observed that the impact of the digital economy on rural revitalization changes with the increase in fiscal expenditure. Specifically, when fiscal expenditure is in the first stage, that is, when the threshold Tt ≤ 0.2818, the impact coefficient of the digital economy on rural revitalization is 0.0568, which is significant at the 5% level. When fiscal expenditure crosses the threshold and enters the second stage, the impact coefficient of the digital economy on rural revitalization increases to 0.1785, which is significant at the 1% level. This shows that the digital economy has always had a positive role in promoting rural revitalization, but there are obvious differences in the intensity of its promotion under different levels of fiscal expenditure.

When fiscal expenditure exceeds a certain threshold, the boosting effect of the digital economy will be significantly enhanced, almost three times that of the first stage. Therefore, when fiscal expenditure levels are high, rural revitalization through the digital economy may achieve better results.

## 4.3 Analysis of moderating effects

(1) The regulatory effect of digital RMB

This article selects the digital renminbi as the regulating variable to further explore its mechanism between the digital economy and rural revitalization, as well as its regulating effect.

Table 4-3 Adjustment model regression results of digital RMB

| | Rural | DR | Rural |
|---|---|---|---|
| constant | 0.178** (6.653) | 5.081** (2.675) | 0.189** (6.845) |
| DEI | 0.001** (1.465) | 0.120*(2.504) | 0.001 (1.052) |
| sample size | 231 | 231 | 231 |
| *R* 2 | 0.307 | 0.376 | 0.324 |
| Adjust *R* 2 | 0.285 | 0.355 | 0.294 |
| *F* value | *F* (3,91)=13.464, *p* =0.000 | *F* (3,91)=18.274, *p* =0.000 | *F* (4,90)=10.772, *p* =0.000 |

Note: The regression coefficients in parentheses are the corresponding t statistics; ***, * *, and * respectively represent 1%, significant at the 5% and 10% levels

Judging from the regression results of the adjusted model in Table 5.4, the fit of the model is also good, and the F value is significant. The regression coefficient of DR is significantly negative, indicating that the issuance of digital RMB has a positive regulatory effect on the development of the digital economy and rural revitalization in Hainan Province. This means that in the context of the central bank's digital currency issuance, it further promotes the development of Hainan Province. The development of the digital economy has indirectly and significantly accelerated the process of rural revitalization.

(2) Regulatory effect of free trade port

Due to the establishment of the Hainan Province Free Trade Port, this article selects time series data to conduct an empirical analysis of the relationship between the digital economy and rural revitalization in the three stages before, during, and after the establishment of the Hainan Province Free Trade Port and examines the role of the free trade port in Hainan Province. After the establishment of Hong Kong, the improvement process of the digital economy and rural revitalization.

Table 4-3 Regression results of the adjustment model of the free trade port

| | Before building | Under Creation | After creation |
|---|---|---|---|
| constant | 0.171**<br>(5.553) | 0.115**<br>(4.453) | 0.214**<br>(6.623) |
| DEI | 0.117***<br>(3.51) | 0.157***<br>(4.41) | 0.313***<br>(4.61) |
| Age | 0.113<br>(0.97) | 0.114<br>(0.97) | 0.114<br>(0.93) |
| CPI | -0.271*<br>(-1.93) | -0.151*<br>(-1.14) | -0.241*<br>(-1.91) |
| Trade | 0.121***<br>(7.52) | 0.111***<br>(7.41) | 0.021***<br>(3.62) |
| GDP | 0.005***<br>(10.03) | 0.004***<br>(10.00) | 0.006***<br>(10.12) |
| Rate1 | 0.313*<br>(1.73) | 0.253*<br>(1.54) | 0.301*<br>(1.44) |
| Rate2 | 0.357*<br>(1.41) | 0.455*<br>(1.57) | 0.424*<br>(1.52) |
| Rate3 | 0.575*<br>(1.77) | 0.474*<br>(1.25) | 0.636*<br>(2.34) |
| rho | -- | -- | -- |
| sigma2_e | -- | -- | -- |
| r2 | 0.836 | 0.837 | 0.842 |
| N | 80 | 80 | 80 |

Note: The regression coefficients in parentheses are the corresponding t statistics; ***, * *, and * respectively represent 1%, significant at the 5% and 10% levels

Judging from the regression results of the adjusted model in Table 5.4, the fit of the model is also good, and the F value is significant. In the comparative analysis of the three stages, the maximum regression coefficient after the establishment of the free trade port was 0.313, which shows that the development of the digital economy in Hainan Province has been significantly improved, which has also further promoted the development of rural revitalization.

## 4.4 GPM optimization strategy prediction

This article studies the impact mechanism of the digital economy and rural revitalization and finds that the development levels of the primary, secondary, and tertiary industries (Rate1, Rate2, and Rate3) all have a significant impact on rural revitalization. Therefore, to further analyze the development path of rural revitalization, this article chooses to construct the development levels of the primary, secondary, and tertiary industries (Rate1, Rate2, Rate3) as the main body of the game and further discusses how the digital economy affects rural revitalization by adjusting the industrial structure. to catalytic effect. The specific parameter symbols and connotation definitions of the GPM model designed this time are shown in Table 4.4:

: Table 4.4 Definition of symbols and related connotations

| symbol | Definition of relevant connotations of symbols |
|---|---|
| $C_g$ | Rate1's investment in Rate2's digital economy to promote rural revitalization |
| $\alpha$ | Rate1 and Rate2 implement subsidies for the digital economy |
| $\beta$ | The income brought by Rate1 to Rate2 |
| $R$ | The losses caused by Rate1 to its economic interests |
| $I$ | Rate1's damage to social credibility |
| $C_f$ | Rate2's investment in implementing the digital economy |
| $C$ | Rate1 is the investment made by Rate1 when Rate2 does not implement the digital economy. |
| $\gamma$ | Rate2's economic and other losses caused by not carrying out digital economic innovation |
| $O$ | Rate2's actual benefits when operating normally using traditional technology |
| $\eta$ | Rate2 takes advantage of the benefits of digital economy innovation |
| $T$ | Rate2 gives Rate3 feedback benefits for participating in the digital economy |
| $C_p$ | Rate3's investment in Rate2 digital economy applications |
| $\varepsilon$ | The benefits Rate3 brings to Rate2's digital economic innovation |
| $\delta$ | Rate3's implementation of the digital economy on Rate2 brings benefits to Rate1 |
| $\Upsilon$ | The losses caused to Rate3 by Rate2's failure to implement digital economic innovation |
| $J$ | The loss of profits caused by Rate3 not carrying out digital economic innovation |

| symbol | Definition of relevant connotations of symbols |
|---|---|
| $v$ | Rate3 does not implement additional purchase investments caused by digital economic innovations |
| $x$ | Rate2's probability of implementing the digital economy |
| $y$ | Rate1The probability of implementing the digital economy |
| $z$ | The probability of Rate3 promoting the Rate2 digital economy |

Table 4.5 Rate1, Rate2, Rate3 three-party betting strategy income function

| situation | strategy combination | Rate1 | Rate2 | Rate3 |
|---|---|---|---|---|
| 1 | (promote, innovate, damage) | $\beta - C_g - \alpha$ | $\eta + \alpha - C_f - T$ | $\varepsilon + \delta + T - C_p$ |
| 2 | (promote, innovate, do not harm) | $\beta - C_g - \alpha$ | $\eta + \alpha - C_f$ | $-J - v$ |
| 3 | (Do not promote, innovate, or damage) | $-R - I$ | $\eta - C_f - T$ | $\varepsilon + \delta + T - C_p$ |
| 4 | (Do not promote, innovate, or damage) | $-R - I$ | $\eta - C_f$ | 0 |
| 5 | (Promotes, does not innovate, harms) | $-C_g - \delta$ | $O - \gamma - C$ | $\delta - C_p - Y$ |
| 6 | (Promote, do not innovate, do not harm) | $-C_g$ | $O - \gamma - C$ | $-J - v$ |
| 7 | (Does not promote, does not innovate, does not harm) | $-R - I$ | $O - \gamma$ | $-C_p - Y$ |
| 8 | (Do not promote, do not innovate, do not harm) | $-R - I$ | $O - \gamma$ | $-J - v$ |

(1) Rate1's replication dynamic equation and asymptotic stability

The expected benefits of Rate1 choosing to promote and not damage the digital economic innovation of Rate2 are represented by $U_{x1}$ and respectively $U_{x2}$. The average expected benefit is $\breve{U}_1$. According to the benefit function obtained in Table 4.5, we can get: The average expected benefit of Rate1 is:

$$\breve{U}_1 = x[yz(\beta - C_g - \alpha) + y(1-z)(\beta - C_g - \alpha) + (1-y)z(-C_g - \delta) + (1-y)(1-z)(-C_g)]$$
$$+(1-x)[yz(-R-I) + y(1-z)(-R-I) + (1-y)z(-R-I) + (1-y)(1-z)(-R-I)]$$

(4.1)

Based on the above-average expected return of Rate1, its replication dynamic equation can be obtained as:

$$E(x) = \frac{dx}{dt} = x(U_{x1} - \breve{U}_1) = x(1-x)[I - C_g + R - y(\alpha - \beta - \delta z) - \delta z] \quad (4.2)$$

Taking ( )the first derivative we get:

$$E'(x) = \frac{dE(x)}{dx} = (1-2x)[I - C_g + R - y(\alpha - \beta - \delta z) - \delta z] \quad (4.3)$$

According to the relevant content of the evolutionarily stable strategy, it can be seen that at that time, the evolutionarily stable strategy could be achieved. $E'(x) < 0$Proposition 1 is obtained: The probability of Rate1 choosing to use the digital economy for high-quality rural development will increase as the probability of Rate1 choosing to implement digital economic innovation without harm to Rate2 increases.

(2) Rate2's replication dynamic equation and asymptotic stability

The expected returns of Rate2 using the digital economy to innovate to promote high-quality rural development and not to innovate are Represented by $\overline{U_2}$and respectively. The average expected return $U_{y2}$ is. According to the income function obtained in Table 3.2, the average expected return of Rate2 is:

$$\breve{U}_2 = y[xz(\eta + \alpha - C_f - T) + x(1-z)(\eta + \alpha - C_f) + (1-x)z(\eta - C_f - T) + (1-x)(1-z)(\eta - C_f)]$$
$$+(1-y)[xz(O - \gamma - C) + x(1-z)(O - \gamma - C) + (1-x)z(O - \gamma) + (1-x)(1-z)(O - \gamma)]$$

(4.4)

From the above average expected return of Rate2, the corresponding replication dynamic equation can be obtained as:

$$E(y) = \frac{dy}{dt} = y(U_{y1} - \breve{U}_2) = y(1-y)[\eta - C_f - O + \gamma + x(\alpha + C) - Tz] \quad (4.5)$$

Looking for $E(y)$the first derivative we can get:

$$E'(y) = \frac{dE(y)}{dy} = (1-2y)\left[\eta - C_f - O + \gamma + x(\alpha + C) - Tz\right] \quad (4.6)$$

According to the relevant content of the evolutionarily stable strategy, it can be seen that at that time, the evolutionarily stable strategy could be achieved. $N'(y) < 0$Proposition 2 is obtained: At that time, $\eta - C_f - O + \gamma + x(\alpha + C) - Tz = 0E(y) = 0$ constant holds, and it is the stable strategy of Rate2 for everything $y$, that is, regardless of whether Rate2 chooses to take advantage of digital economic innovation, Rate2's strategy will not change over time; at that time, according to the evolutionary stability theory, $\eta - C_f - O + \gamma + x(\alpha + C)x - Tz \neq 0$If there is a behavioral strategy $y^*$, $E(y^*) = 0, dE(y)/d(y)|_{j=y^*} < 0$then Rate2 is in a stable state.

(3) Rate3's replication dynamic equation and asymptotic stability

Rate3 choosing to promote Rate2 digital economic innovation and not promoting it are represented by $U_{z1}$and respectively $U_{z2}$. The average expected benefit is. $\overline{\quad}$ According to the income function obtained in Table 3.2, it can be obtained that the average expected benefit of Rate3 choosing to promote Rate2 digital economic innovation is:

$$\bar{U}_3 = z\left[xy\left(\varepsilon + \delta + T - C_p\right) + (1-x)y\left(\varepsilon + \delta + T - C_p\right) + x(1-y)\left(\delta - C_p - Y\right) + (1-x)(1-y)\left(-C_p - Y\right)\right] + (1-z)[xy(-J - v) + x(1-y)(-J - v) + (1-x)(1-y)(-J - v)] \quad (4.7)$$

Based on the above-average expected revenue of the public, we can obtain the corresponding replication dynamic equation as

$$E(z) = \frac{dz}{dt} = z\left(U_{z1} - \bar{U}_3\right) = z(1-z)\left[J - C_p + v - Y + y(\varepsilon - J + \delta + T - v + Y + Jx - \delta x + vx) + \delta x\right] \quad (4.8)$$

Looking for $E(z)$the first derivative we can get:

$$E'(z) = \frac{dE(z)}{dz} = (1-2z)\left[J - C_p + v - Y + y(\varepsilon - J + \delta + T - v + Y + Jx - \delta x + vx) + \delta x\right] \quad (4.9)$$

According to the relevant content of the evolutionarily stable strategy, it can be seen that at that time, the evolutionarily stable strategy could be achieved. $N'(z) < 0$Obtain Proposition 3: The probability of Rate2 choosing to implement digital economic innovation to promote high-quality rural development will increase as the probability of Rate3's actual promotion of Rate2 increases.

Based on this, it can be seen that when the replication dynamic equations of Rate1, Rate2, and Rate3 are all equal to zero, it indicates that the direction of strategic adjustment by the three parties will no longer change, thus making the strategy selection reach a relatively stable equilibrium state.

As a result, at that time, Matlab2021 $E(x) = E(y) = E(z)a$was used to solve and obtain a total of 15 equilibrium points where Rate1 harmed Hainan Province's digital economic innovation promoted by Rate3 and promoted high-quality rural development (Rate2 remained stable), 8 of which met the boundary of evolution $0 \leq x, y, z \leq 1$. The Jacobian matrix of the system can be obtained from the replicated dynamic equations of Rate1, Rate2, and Rate3 respectively, as shown in the formula: $N_1(0,0,0)$, $N_2(1,0,0)$, $N_3(0,1,0)$, $N_4(0,0,1)$, $N_5(1,1,0)$, $N_6(1,0,1)$, $N_7(0,1,1)$, $N_8(1,1,1)$。

$$\begin{bmatrix} (1-2x)\begin{bmatrix} I - C_g + R - \\ y(\alpha - \beta - \delta z) - \delta z \end{bmatrix} & x(1-x)[\beta - \alpha + \delta z] & x(1-x)[\delta y - \delta] \\ y(1-y)[\alpha + C] & (1-2y)\begin{bmatrix} \eta - C_f - O + \gamma \\ +x(\alpha + C) - Tz \end{bmatrix} & y(1-y)[-T] \\ z(1-z)[\delta + y(J - \delta + v] & z(1-z)\begin{bmatrix} \varepsilon - J + \delta + T - v \\ +Y + x(J - Q + v) \end{bmatrix} & (1-2z)\begin{bmatrix} J - C_p + v - Y \\ +y(\varepsilon - J + \delta + T - v \\ +Y + Jx - \delta x + vx) + \delta x \end{bmatrix} \end{bmatrix}$$

Bring the above available equilibrium points back to the Jacobian matrix of the system, and solve to obtain the corresponding equilibrium solution eigenvalues. The specific results are shown in Table 4.6:

Table 4.6 Eigenvalues of each equilibrium solution

| equilibrium point | Eigenvalue 1 | Eigenvalue 2 | Eigenvalue 3 |
|---|---|---|---|
| $E_1(0,0,0)$ | $I - C_g + R(+)$ | $\eta - C_f - O + \gamma(+)$ | $J - C_p + v - Y(+)$ |
| $E_2(1,0,0)$ | $C_g - I - R(-)$ | $J - C_p + \delta + v - Y(+)$ | $\alpha + C - C_f + \eta - O + \gamma(+)$ |
| $E_3(0,1,0)$ | $C_f - \eta + O - \gamma(-)$ | $\varepsilon - C_p + \delta + T(+)$ | $\beta - \alpha - C_g + I + R(-)$ |
| $E_4(0,0,1)$ | $I - C_g - \delta + R(+)$ | $C_p - J - v + Y(+)$ | $\eta - C_f - O - T + \gamma(+)$ |
| $E_5(1,1,0)$ | $\alpha - \beta + C_g - I - R(+)$ | $\varepsilon - C_p + J + \delta + T + v(+)$ | $C_f - C - \alpha - \eta + O - \gamma(-)$ |

| equilibrium point | Eigenvalue 1 | Eigenvalue 2 | Eigenvalue 3 |
|---|---|---|---|
| $E_6(1,0,1)$ | $C_g - I + \delta - R(-)$ | $C_p - J - \delta - v + Y(-)$ | $\alpha + C - C_f + \eta - O - T + \gamma(+)$ |
| $E_7(0,1,1)$ | $C_p - \varepsilon - \delta - T(-)$ | $\beta - \alpha - C_g + I + R(-)$ | $C_f - \eta + O + T - \gamma(-)$ |
| $E_8(1,1,1)$ | $\alpha - \beta + C_g - I - R(-)$ | $C_p - \varepsilon - J - \delta - T - v(-)$ | $C_f - C - \alpha - \eta + O + T - \gamma(-)$ |

Based on this, this article sets the relevant initial conditions of the model based on the relevant game function relationships that have been solved and conducts simulation analysis of a total of 8 strategic selection combinations composed of all behavioral choices in the three parties of Rate1, Rate2, and Rate3.

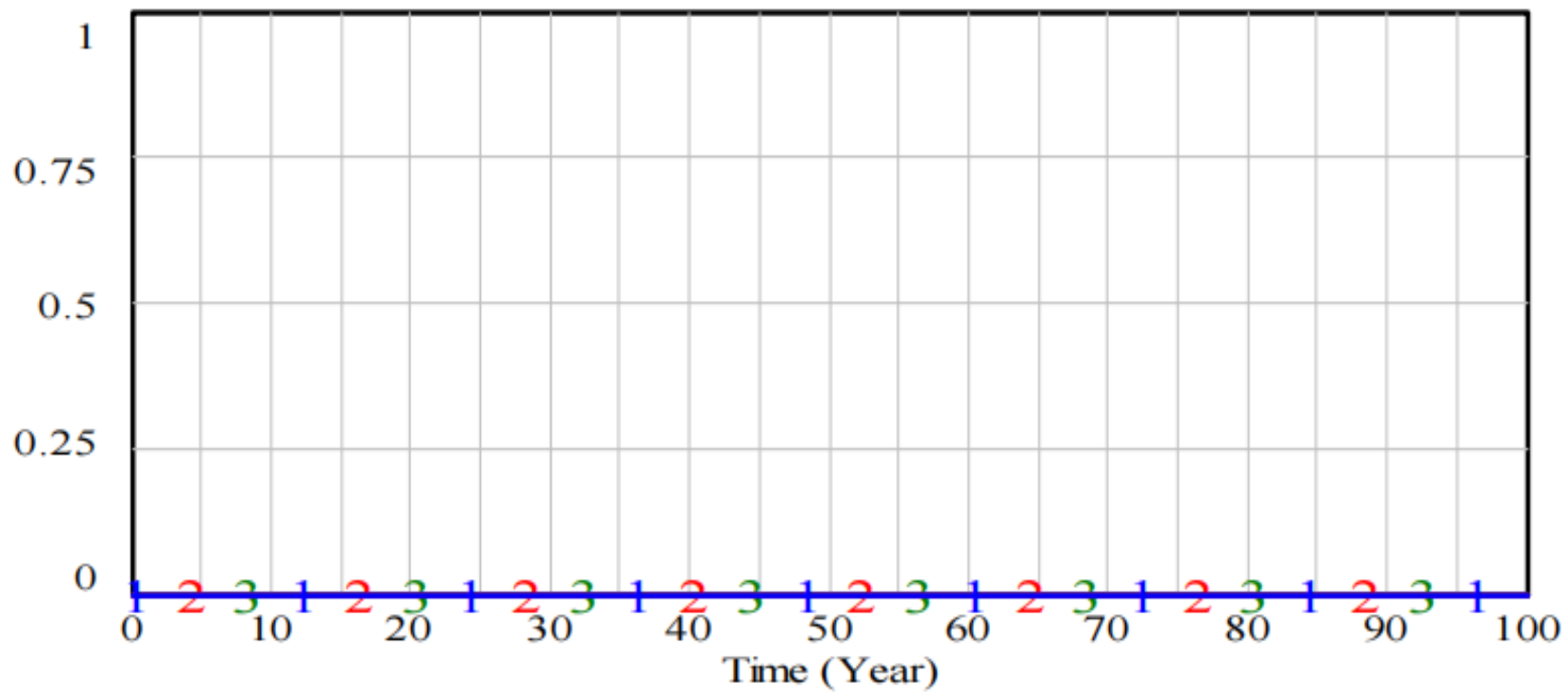

Figure 4.1 Evolutionary game results of pure strategy (0, 0, 0)

By analyzing the initial simulation results of the evolutionary game between the digital economy and high-quality development of rural revitalization, it was found that the three parties were unwilling to take the initiative to change their initial strategic choices in the early stages of the game process. This allowed the three parties to conclude in a short period. Relative stability during the game. However, this equilibrium state is not sustainable and stable. When any of the three parties in the model wants to change its strategic choice, this temporary equilibrium state will change as the strategic choices of the participating entities change. As shown in Figure 4.1, when the strategy combination (0, 0, 0) is substituted into the above SD model, the simulation results show that all parties are unwilling to actively change their initial behavioral choices. When rate1 chooses not to promote, rate2 and rate3 will not actively change their strategies. This is because when rate1 does not harm rate2, rate2 chooses to maintain the original business format to reduce its investment in rural revitalization. Similarly, when rate1 chooses not to promote rate2, rate3 will also choose not to promote rate2 due to the lack of incentives for rate1-related measures and the lack of certain economic benefits.

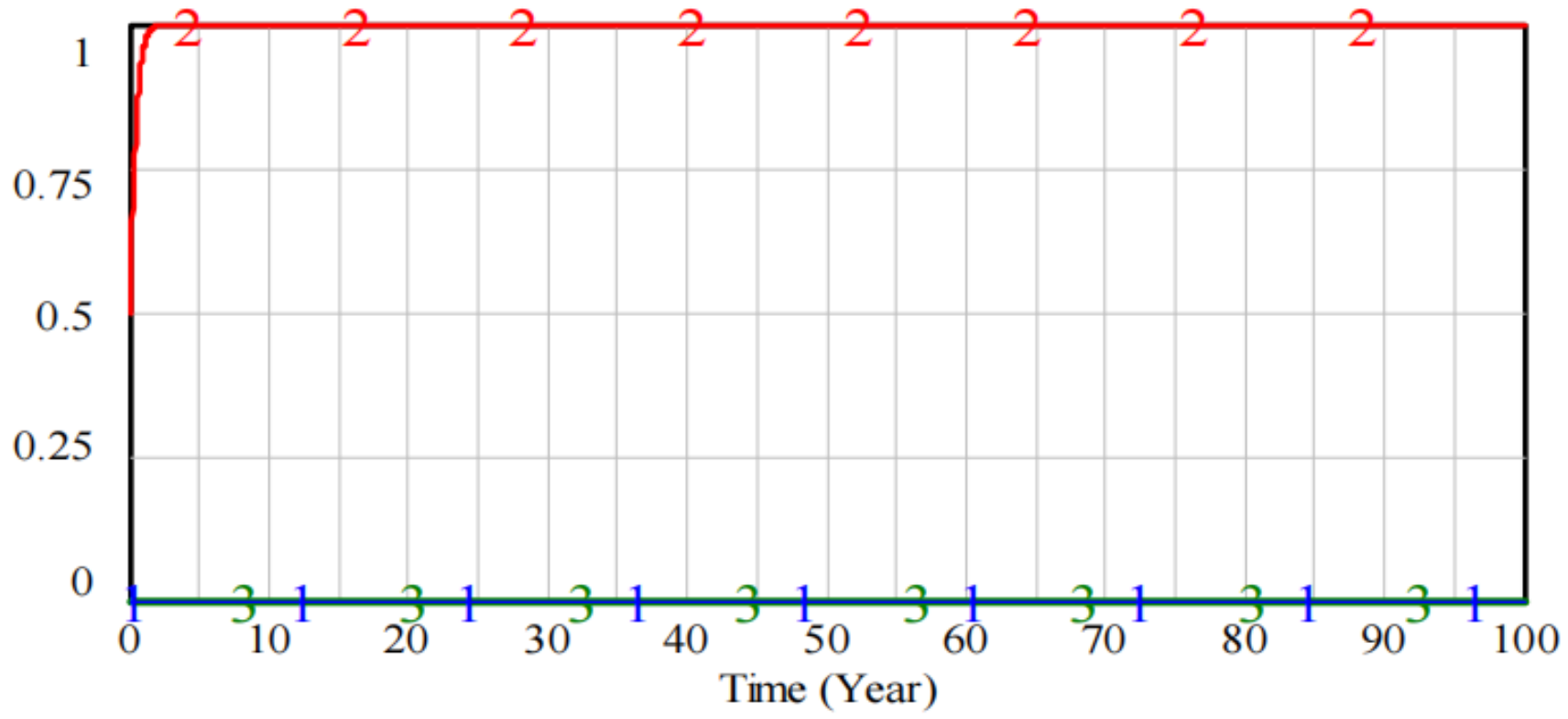

Figure 4.2 Simulation results of rate2 digital economic behavior probability changes

When rate1 and rate3 do not change their initial strategies, the probability of digital innovation behavior of rate2 changes to y=0.01. When the strategy is adjusted, it will be found that the equilibrium solution will evolve from (0, 0, 0) to (0, 1, 0). When rate1 and rate2 do not change the initial strategy, the probability of rate3 changes to z=0.01. If you adjust the strategy, you will find that the equilibrium solution will evolve from (0, 0, 0) to (0, 0, 1). The results of the above two evolutionary games are similar. As shown in Figure 4.2, to increase the economic profits it can obtain, rate2 will choose to independently carry out digital economic innovation to promote the development level of rate2, thereby achieving high-quality development of rural revitalization. Similarly, to safeguard its own economic interests and due rights, rate3 will choose to actively promote rate2.

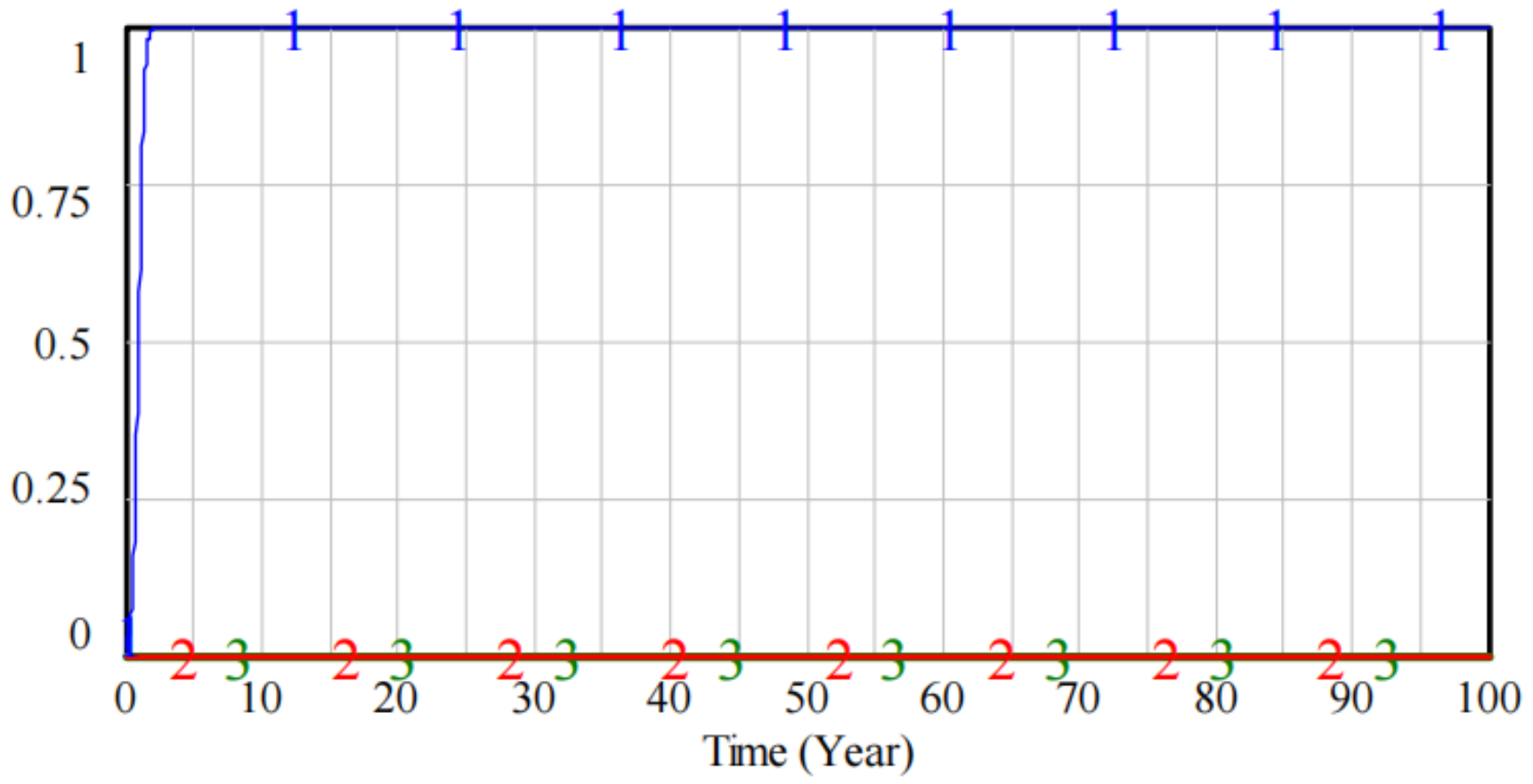

Figure 4.3 Evolutionary game results with rate1 probability change

When rate2 and social rate3 do not change their initial strategy choices, and the probability of rate1 damage is adjusted strategically with x=0.01, as shown in Figure 4.3, the equilibrium solution will rapidly evolve from (0, 0, 0) to (1, 0, 0), when rate1 inflicts damage on rate2, although rate1's damage increases, the benefits gained will be significantly greater than the investment. The damage on rate1 will also make rate2 more willing to actively implement digital economic innovation behaviors and promote the continuous high-quality development of rate2. At the same time, rate1's actions to harm rate2's digital innovation behavior have promoted rate3. Therefore, rate3 is more willing to promote rate2's digital innovation behavior.

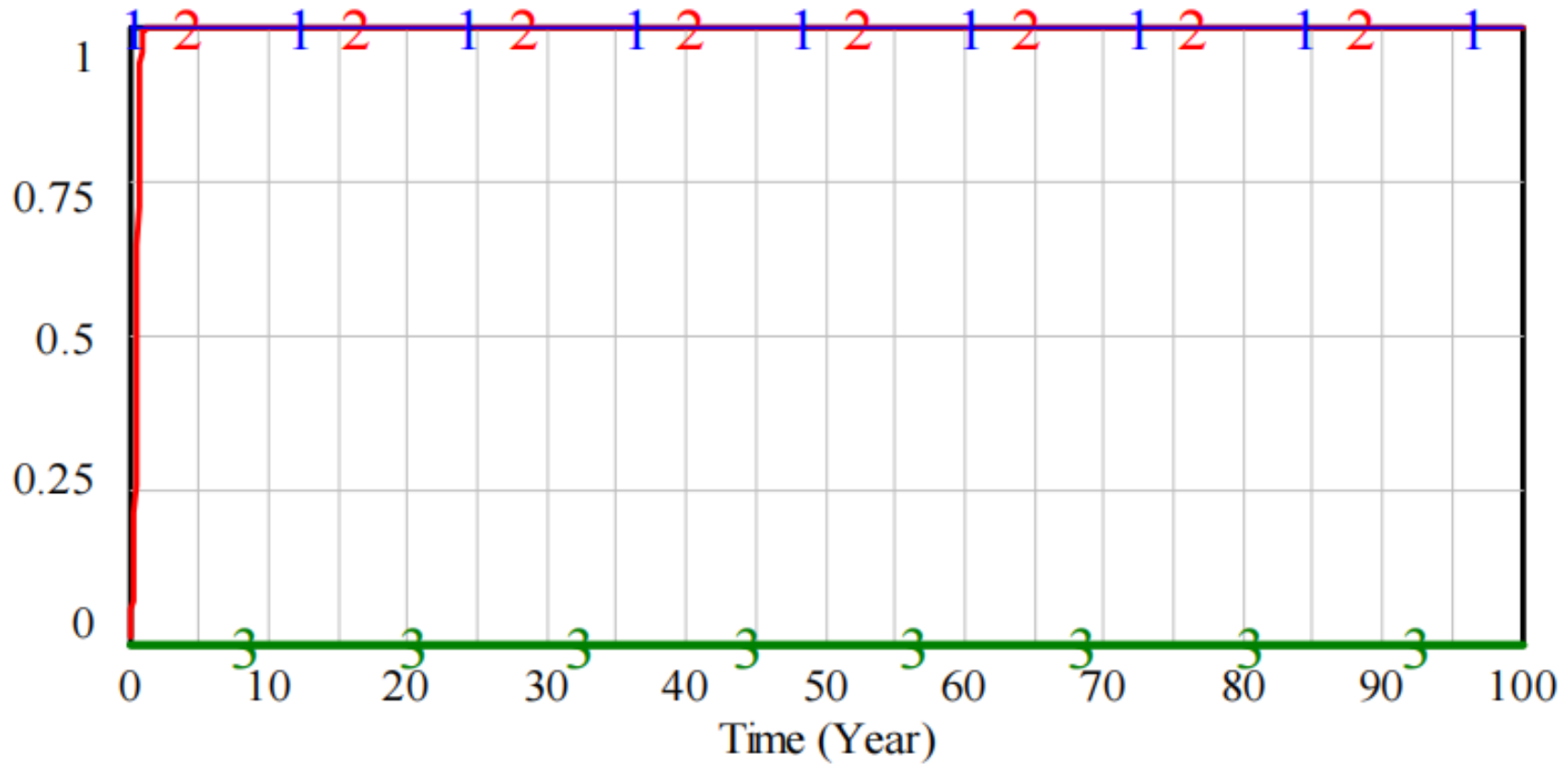

Figure 4.4 Game results of probability change when rate1 implements digital innovation behavior when rate1 is damaged

During the simulation process, it can be found that when rate1 chooses to damage rate2, whether rate2 adjusts its strategy with y=0.01 or rate3 adjusts its strategy with z=0.01, as shown in Figure 4.4, its final evolution strategy will reach 1 The stable state indicates that the damage of rate1 to rate2 will greatly encourage enterprises to proactively implement digital economic innovation behaviors. It also means that the rate2 will vigorously promote rural revitalization and achieve high-quality development under the damage rate1. At the same time, rate3 will be more willing to actively implement promotion through rate1's damage to rate2. When both rate1 and rate3 implement damage and promotion, the industry will use the technological innovation fund subsidies rate2 provides to enterprises to quickly improve its own scientific and technological production level, to gain more trust and support from rate1, and to achieve high-quality rural development.

From the above analysis, it can be seen that when rate1 does not harm rate2, the industry will independently carry out digital economic innovation behaviors to enhance influence and economic interests. Rate3 will also choose to promote rate2 to protect its vested interests, but the promotion and harm are The degree is still lower than when rate1 was involved in the damage; when rate1 participated in the damage, both rate2 and rate3 were more willing to carry out digital innovation and promotion behaviors and assume their responsibilities. When rate1 chooses to harm rate2, rate3 is willing to actively promote rate2's digital innovation behavior to help rate1 better manage rate2 and protect its interests. Therefore, rate1's choice of damage can further encourage enterprises to implement digital economic innovation, encourage rate3 to implement promotional behaviors, improve rate2's sustainable development level, and promote high-quality rural development. On the other hand, when rate3 promotes rate2's digital economic innovation behavior, it can standardize rate2's production behavior to the greatest extent, accelerate the company's faster application of the digital economy to the rural revitalization industry, and promote the technological level of the company. In this case, rate1, rate2, and rate3 all effectively participate in the high-quality development of rural revitalization, bringing the level of rural revitalization development to a higher level and making the overall game reach a stable state of (1, 1, 1), rate1, rate2, and social rate3 can maximize benefits.

## 5 Conclusion

Based on the literature review of the digital economy and rural revitalization, this article conducts an in-depth study of the practice of the digital economy in Hainan Province and draws the following conclusions:

( 1) The digital economy has a significant impact on rural revitalization. The digital economy has brought more development opportunities and employment opportunities to rural areas and improved the living standards and happiness of rural residents by improving agricultural production efficiency, expanding agricultural product sales channels, and promoting the development of emerging industries such as rural tourism and e-commerce.

(2) The development level of the digital economy has a promoting effect on rural revitalization. By constructing an indicator system for the development level of the digital economy and an indicator system for rural revitalization development, and using methods such as panel regression analysis, it was found that the digital economy has a positive promoting effect on rural revitalization. The improvement of the development level of the digital economy will promote the implementation of the rural revitalization strategy.

(3) Digital RMB plays an important regulating role in the digital economy and rural revitalization. As an important part of the digital economy, digital RMB has been widely used in the practice of Hainan Province. By introducing the digital renminbi as an adjusting variable, it is found that it plays an important regulating role between the digital economy and rural revitalization, and can promote the development of the digital economy in promoting rural revitalization.

(4) The integration of the digital economy and rural revitalization requires the joint efforts of the government, enterprises, and society. When implementing the rural revitalization strategy, the government should strengthen the construction of digital infrastructure, improve relevant policies and regulations, and strengthen talent training; enterprises should actively participate in the development of the digital economy and promote the digital transformation of industries; all sectors of society should jointly pay attention to and support digital Economic development, improving digital literacy and skill levels.

(5) In the prediction of the optimal strategy for rural revitalization based on the development levels of the primary, secondary, and tertiary industries (Rate1, Rate2, and Rate3), it was found that the interaction and cooperation between the primary, secondary, and tertiary industries have an important impact on rural revitalization. The high-quality development of rate1 is crucial. rate1 can further encourage enterprises to implement digital economic innovation, encourage rate3 to implement promotion behaviors, and improve the level of sustainable development of rate2. When rate3 promotes rate2 digital economic innovation behaviors, it can standardize rate2 production behaviors to the greatest extent. , accelerate the faster application of the digital economy to the rural revitalization industry, and promote the technological level of enterprises. Through effective cooperation and coordination, these industries can jointly promote sustainable rural development and maximize overall benefits.

The digital economy is an important force in promoting rural revitalization and is of great significance to agricultural and rural modernization. Hainan Province should make full use of the advantages of the digital economy, strengthen policy guidance and planning support, promote the in-depth integration of digital technology with agriculture and rural industries, and promote the overall revitalization of rural areas. At the same time, we should pay attention to the challenges and risks of the development of the digital economy, strengthen supervision and preventive measures, and ensure the healthy development of the digital economy.